\documentclass[preprint,superscriptaddress,nofootinbib]{revtex4-2}

\usepackage[T1]{fontenc}
\usepackage[utf8]{inputenc}
\usepackage{lmodern}

\usepackage{graphicx,amssymb,amsmath,amsfonts}
\usepackage{amsthm}
\usepackage[english]{babel}
\usepackage{hyperref}
\usepackage{cleveref}

\usepackage{subcaption}
\usepackage{xcolor}
\usepackage{mathrsfs}
\usepackage{slashed} 

\newcommand{\p}{{\partial}}
\newcommand{\w}{{\omega}}

\newcommand{\ep}{\epsilon}

\newcommand{\zb}{{\bar{z}}}

\newcommand{\mH}{\mathcal{H}}
\newcommand{\mI}{\mathcal{I}}

\newcommand{\mQ}{\mathcal{Q}}

\let\O\relax
\newcommand{\O}{O}

\newcommand{\scri}{\mI}

\let\tilde\relax
\newcommand{\tilde}{\widetilde}

\DeclareMathOperator{\tr}{tr}
\DeclareMathOperator{\diag}{diag}

\newcommand{\s}{{\mH^+_-}}

\let\H\relax
\newcommand{\H}{{\mH^+}}

\newcommand{\QEfull}[1][f]{{Q_{#1}^\H}}
\newcommand{\QMfull}[1][f]{{\tilde Q_{#1}^\H}}
\newcommand{\QE}[1][f]{{\slashed\delta\QEfull[#1]}}
\newcommand{\QM}[1][f]{{\slashed\delta\QMfull[#1]}}
\newcommand{\iQE}[1][f]{{\delta\QEfull[#1]}}
\newcommand{\iQM}[1][f]{{\delta\QMfull[#1]}}

\newcommand{\eQE}[1][\l]{{\mQ_{#1}^\H}}
\newcommand{\eQM}[1][\l]{{\tilde\mQ_{#1}^\H}}

\newcommand{\x}{\Theta}

\begin{document}

\title{Holography from Singular Supertranslations on a Black Hole Horizon}

\author{Ratindranath Akhoury}
\email{akhoury@umich.edu}
\affiliation{Leinweber Center for Theoretical Physics, Randall Laboratory of Physics, Department of Physics, University of Michigan, Ann Arbor, MI 48109, USA.}

\author{Sangmin Choi}
\email{sangmin.choi@polytechnique.edu}
\affiliation{CPHT, CNRS, \'Ecole Polytechnique, Institut Polytechnique de Paris, F-91128 Palaiseau, France.}

\author{Malcolm J. Perry}
\email{malcolm@damtp.cam.ac.uk}
\affiliation{School of Physics and Astronomy, Queen Mary University of London, Mile End Road, London E1 4NS, UK.}
\affiliation{DAMTP, Centre for Mathematical Sciences, Cambridge University, Wilberforce Road, Cambridge CB3 OWA, UK.}
\affiliation{Trinity College, Cambridge, CB2 1TQ, UK.}

\begin{abstract}
We investigate the standard and dual Bondi-Metzner-Sachs (BMS) supertranslation	
generators on a black hole horizon and draw some conclusions about black hole physics.	
Recently, it has been shown that in addition to conventional BMS supertranslation symmetries, 	
there exists an additional infinite set of magnetic	
asymptotic symmetries, dual BMS supertranslations, again parametrized by a function on the two-sphere.	
We show that the Dirac bracket between these generators exhibits an anomalous central	
term when one parameter function exhibits a singularity in the complex	
stereographical coordinates on the sphere. In order to preserve general coordinate invariance, we demonstrate that this central term can be removed by postulating a holographic gravitational Chern-Simons theory on the horizon.	
This indicates that for an anomaly-free theory of quantum gravity in the presence of	
a black hole, one should include a boundary theory on
the horizon.
\end{abstract}

\hfill CPHT-RR034.052022

\maketitle

\section{Introduction}\label{sec:intro}
The recent interest in asymptotic symmetries \cite{Hawking:2016msc,Hawking:2016sgy} is motivated in large part by the applications to black hole physics. The asymptotic symmetries of particular interest in this Letter are the Bondi-Metzner-Sachs (BMS) supertranslations and the dual supertranslations. While the supertranslations are associated with a certain class of diffeomorphisms, parametrized by a function on a two-sphere, no such association is known for the dual ones. However, in refs$.$ \cite{Kol:2019nkc,Godazgar:2019dkh}, it was noted that the dual symmetries are contained in a complexification of the usual supertranslation. The action of the complexified charge on phase space was also investigated there, indicating that this situation is completely analogous to what happens in electromagnetism, \cite{Strominger:2015bla}. This analogy with electromagnetism is further explored in this Letter by considering the algebra of the dual and standard supertranslation charges (magnetic and electric charges in electromagnetism) for a
	parameter function with singularities in  terms of complex stereographical coordinates on the sphere.
There are several physical motivations for considering singular parameter functions.
In electromagnetism, Dirac-string-like configurations in the bulk correspond to singular gauge transformations,
	and thus in gravity it is natural to expect that singular supertranslations correspond to the gravitational analog of such configurations.
Moreover, singular supertranslations arise naturally in the BMS algebra of supertranslation and superrotation charges \cite{Barnich:2011mi}.
While a proper understanding of BMS superrotations is a subject of ongoing research,
	considering singularities in the supertranslation parameter would be one of the first steps in this direction.
Singular gauge transformations in electromagnetism were considered in \cite{Hosseinzadeh:2018dkh,Freidel:2018fsk}, and it was found that the algebra of the electric and magnetic charges at null infinity contained a central term.
Since asymptotic symmetry charges map one physical configuration to another, one expects the charge algebra to be closed.
Thus, a central term appearing in the algebra signals the existence of an anomaly.

In this Letter, we consider pure gravity without matter and show that an analogous central term arises in the algebra of standard and dual supertranslation charges at the future black hole horizon even if there are no bulk Newman-Unti-Tamburino charges. We further find that this anomaly in the charge algebra may be canceled by including a holographic Chern-Simons theory on the horizon of the black hole. We give an outline discussion of the electromagnetic case first and construct the Chern-Simons theory here as well, so as to bring out the analogies and differences between the two theories. We find the Chern-Simons theory for electromagnetism to have gauge group
 $\mathrm{U}(1)\otimes \mathrm{U}(1)$, in agreement with a similar result on future null infinity in \cite{Freidel:2018fsk}. For the case of gravitation, we find the Chern-Simons theory to have gauge group $\mathrm{SL}(2,\mathbb C)$ on the horizon. The implications of this for a quantum theory of gravity, in particular, the relevance of the states of the Chern-Simons theory to black hole physics, is beyond the scope of this Letter but will be treated elsewhere.
 
 In section \ref{sec:bms}, we review the analog of both the standard and dual supertranslations on the black hole  horizon in Bondi gauge. In section \ref{sec:charges}, we introduce the charges associated with
 the diffeomorphism symmetries. In parallel, we also discuss the dual, magnetic, counterpart of the
 diffeomorphism symmetries.  In section \ref{sec:algebra}, we continue the discussion of charges and allow for the possibility
 that there could be singularities in the supertranslations. We examine in detail the case of 
 the supertranslation generator having  a simple pole when expressed in the usual complex coordinates on 
 the two-sphere of the horizon and show that the algebra of electric (standard) and 
 magnetic (dual) supertranslation charges is anomalous. The central charge is explicitly calculated here.
 In section \ref{sec:em},
 we examine electromagnetic soft hair on the black hole horizon and show that in this case, also, there is an anomaly in the charge algebra when one has both electric and magnetic transformations. We show that this anomaly can be canceled by supposing that the horizon has a Chern-Simons theory living on it. In section \ref{sec:cs}, we repeat this analysis for the gravitational case and find that the Chern-Simons theory that does the job here was formulated earlier in \cite{Witten:1988hc}. 
 We conclude with a brief discussion of our results.

\section{BMS transformations}\label{sec:bms}

We begin by reviewing the notion of supertranslations on the future horizon of the Schwarzschild black hole. The Schwarzschild metric in terms of the advanced Eddington-Finkelstein coordinate $v$, and the stereographic coordinates $z^A$ ($A=1,2$) which parametrize a unit  2-sphere with metric $\gamma_{AB}$ is given by,
\begin{align}
	ds^2 = g_{ab}dx^adx^b = -\left(1-\frac{2M}{r}\right) dv^2 + 2dvdr + r^2 \gamma_{AB} dz^A dz^B.
\end{align}
We will work in the Bondi gauge, where $g_{rr} = 0$, $g_{rA}=0$, $\p_r \det\frac{g_{AB}}{r^2}=0$ where
$g_{AB}$ is the metric on the two-sphere.
The diffeomorphisms that respect the Bondi gauge conditions are generated by the vector field \cite{Hawking:2016msc,Hawking:2016sgy},
\begin{align}
	\xi^a\p_a =
	X \p_v - \frac{1}{2}\left(rD_A X^A + D^2 X\right) \p_r + \left(X^A + \frac{1}{r}D^A X\right)\p_A
	,
	\label{xi}
\end{align}
where, $D_A$ is the covariant derivative with respect to $\gamma_{AB}$ and the $A,B,\ldots$ indices are raised
and lowered using the metric on the unit two-sphere $\gamma_{AB}$.
Supertranslations are generated by $X=f$ and $X^A=0$, while superrotations are generated by $X=\frac{u}{2}D_AY^A$, $X^A=Y^A$. Here $f$ parametrizes the diffeomorphism on a two-sphere (either at null infinity or the horizon). 
In this Letter, we will be interested in the charge associated with supertranslations and the related dual supertranslations. 

The dual gravitational charges are given by twisted fields \cite{Godazgar:2019dkh,Choi:2019sjs} defined through the Levi-Civita tensor $\epsilon _{AB}$ on the two-sphere. Unlike the supertranslation charge, they do not generate diffeomorphisms and may be derived, as recently shown in \cite{Godazgar:2020kqd,Godazgar:2020gqd}, by introducing the topological Holst term to the Einstein action. The most convenient way to view the dual symmetries is through a complexification of supertranslation charge \cite{Godazgar:2019dkh}. It was shown in this last reference that the complexified charge acts on the phase space in a manner consistent with what one would expect for the analogous case of electric and magnetic charges in electromagnetism \cite{Strominger:2015bla}. Specifically, the complex charge on future null infinity is given by \cite{Godazgar:2019dkh}
\begin{align}
\textbf{Q}^+_f = -\frac{1}{16\pi}\int_Sd\Omega(-4fm_B + fD^2_zC^{zz} + D_{{\bar{z}}}fD_{{\bar{z}}}C^{{\bar{z}}{\bar{z}}}),
\end{align}
where, $S$ is the two-sphere at future null infinity, $rC_{AB}=\delta g_{AB}$, $\delta g_{AB}$ is the variation of the metric $g_{AB}$ and $m_B$ the Bondi mass aspect. They further find that its action on phase space may be represented by,
\begin{align}
\{\textbf{Q}^+_f, C_{zz}(u,z,{\bar{z}})\} &= f\partial_uC_{zz} \\
\{\textbf{Q}^+_f, C_{{\bar{z}}{\bar{z}}}(u,z,{\bar{z}})\} &= f\partial_uC_{{\bar{z}}{\bar{z}}} - 2D^2_{{\bar{z}}}f.
\end{align}
Thus, $\textbf{Q}^+_f$ generates time translation on the mode, $C_{zz}$ and a supertranslation on the other mode $C_{{\bar{z}}{\bar{z}}}$ in complete analogy with the electromagnetic case discussed in \cite{Strominger:2015bla}.
\section{Horizon charges}\label{sec:charges}
Let us now construct charges that generate the standard and dual BMS supertranslations on the horizon.
Consider a spacelike hypersurface $\Sigma$ that extends from a section of $\scri^+$ to a section of $\mH^+$.
A charge $Q_f^\Sigma$ associated to $\Sigma$ then splits into two parts,
\begin{align}
	Q_f^\Sigma = Q_f^{\H} + Q_f^{\mI^+}
	,
\end{align}
where $Q_f^{\H}$ is defined on a section on the future horizon $\mH^+$ and $Q_f^{\mI^+}$ is defined on a section on the future null infinity $\mI^+$.
We refer to the horizon contribution $Q_f^\H$ as the horizon charge.

The horizon charge can be written as an integral on $\H$ so long as the contribution from the future boundary $\mH^+_+$ vanishes.
In this Letter, we take the viewpoint that the black hole ultimately evaporates,
	and hence this contribution is negligible.
If a horizon has a future end-point, in classical general relativity this point must be singular.
Following common practice, we assume that this is not an issue and that the quantum theory will ultimately resolve this problem.
The horizon charge can therefore be written either as a 3-dimensional integral of a total $v$-derivative over $\H$, or a 2-dimensional integral over $\mH^+_-$

To each vector field $\xi$ that does not vanish on $\Sigma$, there exists a diffeomorphism (electric) horizon charge $Q_f^\H$ associated to the diffeomorphism generated by $\xi$,
	as well as a dual (magnetic) horizon charge $\tilde Q_f^\H$ that originates from the Holst action \cite{Godazgar:2020kqd}.
Variations of such charges can be computed using the covariant phase space formalism in the first-order formulation of gravity \cite{Godazgar:2020kqd,Godazgar:2020gqd}.
Taking the vector field $\xi$ to be the supertranslation generator \eqref{xi}, we obtain the standard and dual supertranslation horizon charges,
\begin{align}
	\slashed \delta Q^\H_f 
	&=
		\frac{1}{16\pi M}\int_\H dv\,d^2\x \sqrt \gamma f(\x) D^A D^B \sigma_{AB}
	,\\
	\slashed \delta \tilde Q^\H_f
	&=
		\frac{-1}{32\pi M}\int_{\p\H} d^2\x\sqrt\gamma
			(D^B f) \ep_A{}^CD^Ah_{BC}
	,
\end{align}
where $h_{AB}=\delta g_{AB}$ is the variation of the metric, and $\sigma_{AB}=\frac{1}{2}\p_v h_{AB}$ is its conjugate momentum.
It is possible that these variations are non-integrable, and we emphasize this point with the notation $\slashed\delta$,
	in contrast to integrable variations denoted by $\delta$.

\section{Charge algebra}\label{sec:algebra}

In this section, we demonstrate that the Dirac bracket algebra between the standard and dual supertranslation charges on the horizon
	exhibits a central term when the parameter function exhibits a singularity.
To demonstrate this, let us consider a supertranslation with a pole $f(z,\zb)=\frac{1}{z-w}$ and a dual supertranslation with a smooth function $f'$.\footnote{We work with a simple pole for explicit computation,
	but a similar line of argument can be carried out for other functions with singularities, such as logarithms.}
Then, the dual charge for smooth $f'$ becomes integrable $\QM[f']=\iQM[f']$,
	but the supertranslation charge acquires a non-integrable piece,
\begin{align}
	\QE[f=\frac{1}{z-w}] &= \iQE[f=\frac{1}{z-w}] - \frac{1}{4M}\int_{-\infty}^\infty dv D_z [D^2-1]^{-1}D^BD^A \sigma_{AB}\bigg|_{z=w}
	.
\end{align}
It is straightforward to check that the non-integrable piece has vanishing Dirac brackets with all charges,
	and therefore can be ignored for our computation of brackets.
Natural definitions of the integrable variations are given by
\begin{align}
	\iQE
	&\equiv
		\frac{1}{16\pi M}\int_\H dv\,d^2\x \sqrt \gamma\, (D^BD^Af) \sigma_{AB}
	\label{iQE}
	,\\
	\iQM[f']
	&\equiv
		\frac{-1}{32\pi M}\int_\s d^2\x \sqrt \gamma\, (D^BD^Af') \ep_A{}^Ch_{BC}
	\label{iQM}
	.
\end{align}
It is worth noting that $\iQM$ is related to $\iQE$ by the twisting $h_{AB} \to \ep_A{}^C h_{CB}$ \cite{Godazgar:2018dvh,Godazgar:2019dkh,Choi:2019sjs}.
These variations are first-order perturbations around the Schwarzschild background.
Thus, one can imagine that a full (integrated) charges $Q$ has the expansion
\begin{align}
	Q
	&=
		Q_0
		+ \delta Q
		+ \O(h^2)
	.
\end{align}
Here $Q_0$ is the charge evaluated on the Schwarzschild metric, which is a constant and hence does not carry degrees of freedom.
This implies that at leading order in $h$, the Dirac bracket between the full charges $\QEfull$ and $\QMfull[f']$ is
\begin{align}
	\{
		\QEfull,
		\QMfull[f']
	\}
	&=
		\{
			\iQE,
			\iQM[f']
		\}
		+ \O(h)
	.
\end{align}
Since both $\iQE$ and $\iQM[f']$ are linear in $h_{AB}$ and $\sigma_{AB}$ respectively,
	their Dirac bracket is a constant.
This implies that the bracket $\{\iQE,\iQM[f']\}$ contains information about the central term of the full charge algebra.
By straightforward computation, one finds that
\begin{align}
	\left\{\iQE[f=\frac{1}{z-w}], \iQM[f']\right\}
	&=
		\frac{-i}{4} (D^z D_z^2 f')\bigg|_{z=w}
		.
\end{align}
One observes the appearance of a central term in the algebra.

\section{Electromagnetism}\label{sec:em}

In the previous section, we have shown that the presence of a singularity in the supertranslation parameter function leads to a central term in the charge algebra.
In this section, we review similar results for electromagnetism, and show that such central term can be removed by adding a Chern-Simons theory on the horizon.
We refer the reader to \cite{Hosseinzadeh:2018dkh,Freidel:2018fsk} for related discussions on $\mI^+$.

In electromagnetism, the asymptotic symmetries are the electric and magnetic
	large gauge transformations.
The soft horizon charges are given by the expressions
\begin{align}
	\eQE
	&=
		\int_{\H} d\psi\wedge *F
	,\qquad
	\eQM[\sigma]
	=
		\int_{\H} d\sigma\wedge F
	,
\end{align}
where $\psi=\psi(z, {\bar{z}})$ and $\sigma=\sigma(z, {\bar{z}})$ are functions on the sphere.
It is straightforward to see that they satisfy the algebra
\begin{align}
	\{\eQE,\eQM[\sigma]\} &= \int_{\p\H} d\psi\wedge d\sigma
	,\\
	\{\eQE,\eQE[\sigma]\} &= 0
	,\quad
	\{\eQM,\eQM[\sigma]\} = 0
	.
\end{align}
One observes that singularities in $\psi$ (or $\sigma$) lead to central terms in the algebra.

Central terms in a symmetry algebra imply the existence of anomalies.
One way to remove such a term is to add a boundary theory on $\H$ that has an asymptotic charge algebra with a central term canceling it.
For this purpose, let us consider a $\mathrm{U}(1)\otimes \mathrm{U}(1)$ Chern-Simons theory on $\H$,
\begin{align}
	S = \int_{\H} a\wedge d\tilde a
	,
\end{align}
where $a$ and $\tilde a$ are the electric and magnetic $\mathrm{U}(1)$ gauge fields.
Under an electric large gauge transformation (LGT), $\delta a=d\phi$ and $\delta \tilde a=0$,
	and under a magnetic one $\delta a=0$ and $\delta\tilde a=d\tilde\phi$.
They are generated by the charges
\begin{align}
	\delta \mQ_\phi
	&=
		-\int_{\p\H}
			d\phi\wedge \delta \tilde a
	,\\
	\delta \tilde \mQ_{\phi}
	&=
			\int_{\p\H}
			\delta a\wedge d\phi
	,
\end{align}
respectively.
The charge algebra is
\begin{align}
	\{ \mQ_{\phi},\tilde\mQ_{\varphi}\}
	&=
		-\int_{\p\H} d\phi\wedge d\varphi
	,\\
	\qquad
	\{ \mQ_{\phi},\mQ_{\varphi}\}
	&=
	\{ \tilde\mQ_{\phi},\tilde\mQ_{\varphi}\}
	=
		0
	.
\end{align}
Therefore, one finds the algebra to be exactly parallel to that of standard and dual LGT charges on the horizon.
This tells us that putting a $\mathrm{U}(1)\otimes \mathrm{U}(1)$ Chern-Simons theory
	 on the horizon, we can get rid of the central term in the standard and dual LGT algebra.

\section{Gravitational Chern-Simons theory}\label{sec:cs}
In this section, we will follow Witten's approach to gravity in three dimensions \cite{Witten:1988hc}
	and consider a gravitational Chern-Simons theory on the horizon. The algebra of the charges in this theory is calculated and the parameters chosen such that the central term here cancels the one found in section \ref{sec:algebra}.
In three dimensions, we use $i,j,\ldots$ for spacetime indices and $a,b,\ldots$ for tangent space.
The action is given in the canonical form by,
\begin{equation}
I_{CS} = \frac{k}{4\pi}\int \tr\ \bigl(A\wedge dA + \tfrac{2}{3} A \wedge A \wedge A\bigr).
\end{equation}
Equations of motion fix the gauge group to be ${\bf G}=\mathrm{SL}(2,\mathbb C)$.
The gauge field decomposes as $A_i = e^a_i P_a + \w^a_i J_a$, where the generators $P_a$ and $J_a$ satisfy the commutators
\begin{align}
	[J_a,J_b] = \ep_{abc}J^c
	,\quad
	[J_a,P_b] = \ep_{abc}P^c
	,\quad
	[P_a,P_b] = \lambda \ep_{abc}J^c
	,
\end{align}
where the Cartan metric $\eta_{ab}=\diag(-++)$ is used to lower and raise tangent space indices,
	and $\lambda=-\frac{1}{4M^2}$ is a negative constant.
As the names suggest, Witten found that this Chern-Simons theory is equivalent to a first-order Einstein theory
	after identifying $e^a_i$ and $\w^a_i$ as the vielbein and the spin connection.

There exist two Killing forms and therefore one can write two actions in terms of $e^a$ and $\w^a$.
The first Killing form is $\langle J_a,P_b \rangle = \eta_{ab}$, with all other components vanishing.
It leads to the \textit{electric} action
\begin{equation}
I_\text{electric}  = \frac{k}{2\pi}\int_{\H} 2e^a\wedge d\omega_a + \epsilon_{abc}  e^a\wedge \omega^a\wedge \omega^c
+\, \tfrac{1}{3}\lambda \epsilon_{abc} e^a\wedge e^b\wedge e^c \label{eq:actA}
.
\end{equation}
The second Killing form has non-vanishing components $\langle J_a,J_b \rangle = \eta_{ab}$, 
$\langle P_a,P_b \rangle = \lambda\eta_{ab}$,
	and it leads to the \textit{magnetic} action,
\begin{equation}
I_\text{magnetic} = \frac{\tilde k}{\pi}\int_\H \omega^a\wedge d\omega_a + \tfrac{1}{3}\epsilon_{abc}\omega^a \wedge \omega^b
\wedge \omega^c + \lambda e^a \wedge de_a + \lambda\epsilon_{abc}\omega^a\wedge e^b \wedge e^c. 
\label{eq:actB}
\end{equation}
$\tilde{k}$ is arbitrary here but we will see later that it can be expressed in terms of $k$ by demanding that the complexified charges form a closed algebra in the absence of singularities in the function parametrizing the diffeomorphisms. 
Drawing analogy to four-dimensional gravity, the electric action corresponds to the Einstein-Hilbert action whereas
	the magnetic action corresponds to the Holst action \cite{Godazgar:2020kqd}.
As in four dimensions, we take the electric action to be our gravitational Chern-Simons theory,
	and use the magnetic action only to derive the dual charge.

There are two types of gauge transformations, one labeled by a tangent-space vector $\rho^a$ and the other by a second vector $\tau^a$.
The fields transform as $\delta e_i^a = -\partial_i\rho^a - \epsilon^{abc}\omega_{ib} \rho_c$ and $\delta\omega_i^a = -\lambda\epsilon^{abc}e_{i\,b}\rho_c$ under $\rho$,
	and $\delta e_i^a = -\epsilon^{abc}e_{ib} \tau_c$ and $\delta \omega_i^a = -\partial_i\tau^a - \epsilon^{abc}\omega_{ib} \tau_c$ under $\tau$.
The $\tau$-transformations correspond to Lorentz transformations on the tangent space,
	whereas the $\rho$-transformations correspond to diffeomorphisms $\rho^a = \imath_v e^a$ generated by the spacetime vector field $v$ up to a compensating Lorentz transformation.

Using the covariant phase space formalism, one finds the electric and magnetic charges associated with the gauge transformations to be \cite{ACP}
\begin{align}
Q^E_{\rho,\tau} &= -\frac{k}{\pi} \int_{\partial \Sigma} \ \tau_ae^a+\rho_a\omega^a 
,\\
Q^M_{\rho,\tau} &= -\frac{2\tilde k}{\pi} \int_{\partial \Sigma} \tau_a\omega^a + \lambda\rho_ae^a
.
\end{align}
The domain of integration $\partial \Sigma$ is the collection of closed loops around the singularities.\footnote{
This implies that the charges vanish for smooth parameter functions.
However, we are interested in the variation of these charges under the action of another charge with singularities, which may not be zero.
}
Having computed the magnetic charge, we dispose of the magnetic action and work solely with the electric action.
The symplectic structure of the electric theory dictates that the only non-vanishing Dirac bracket of the theory is
\begin{align}
	\{e^a_z(z,\zb),w^b_\zb(z',\zb')\} = -\frac{i\pi}{k}\eta^{ab}\delta^2(z-z')
	,
\end{align}
where $z,\zb$ are stereographic coordinates of the unit two-sphere.
Using the brackets, one finds the following algebra of electric and magnetic charges \cite{ACP},
\begin{align}
	\{Q^E_{\tau,\rho},Q^E_{\tau',\rho'}\}
	&=
		Q^E_{\tau'',\rho''}
		- \frac{k}{\pi}\int_{\p\Sigma} \left(
			\rho^a d\tau'_a
			+ \tau^a d\rho'_a
		\right)
	,\\
	\{Q^E_{\tau,\rho},Q^M_{\tau',\rho'}\}
	&=
	Q^M_{\tau'',\rho''}
		- \frac{2\tilde k}{\pi}\int_{\p\Sigma}\left(
			\tau_a d\tau'^a
			+ \lambda \rho_a d\rho'^a
		\right)
	,\\
	\{Q^M_{\tau,\rho},Q^M_{\tau',\rho'}\}
	&=
		4\lambda\frac{\tilde k^2}{k^2} Q^E_{\tau'',\rho''}	
		- \frac{4\lambda\tilde k^2}{\pi k}\int_{\p\Sigma} \left(
			\rho^a d\tau'_a
			+ \tau^a d\rho'_a
		\right)
	.
\end{align}
The composition is given by $\tau''^a=\ep^{abc}(\tau'_b\tau_c+\lambda\rho'_b\rho_c)$ and $\rho''^a=\ep^{abc}(\tau'_b \rho_c - \tau_b \rho'_c)$.
We demand that the central term of this algebra cancels the central term observed in the supertranslation algebra on the Schwarzschild horizon which we computed for $f=\frac{1}{z-w}$.
Since supertranslation is a diffeomorphism, it acts as a gauge transformation $\rho^a = \iota_v e^a$ on the horizon Chern-Simons theory with $v=f\p_v+\frac{1}{2M}D^Af\p_A$,
	up to a Lorentz transformation.
We define the compensating Lorentz transformation of supertranslation to be
\begin{align}
	\tau^0 = \frac{(D^2+2)f}{8(2\tilde k)^{1/2}}
	,\quad
	\tau^1 = i\sqrt\lambda\rho^2
	,\quad
	\tau^2 = -i\sqrt\lambda\rho^1
	.
\end{align}
Note that all components are real since $\lambda$ is negative.
Then algebra of charges with $f=\frac{1}{z-w}$, becomes
\begin{align}
	\{Q^E_{\tau,\rho},Q^E_{\tau',\rho'}\}
	&=
		Q^E_{\tau'',\rho''}
	,\\
	\{Q^E_{\tau,\rho},Q^M_{\tau',\rho'}\}
	&=
		 Q^M_{\tau'',\rho''}
		+ \left.\frac{i}{4} (D^z D_z^2 f')\right|_{z=w}
	,\\
	\{Q^M_{\tau,\rho},Q^M_{\tau',\rho'}\}
	&=
		4\lambda\frac{\tilde k^2}{k^2} Q^E_{\tau'',\rho''}
	.
\end{align}
The central term obtained here is exactly of the form to cancel the one obtained in section \ref{sec:algebra}. Note that $f'$ in the above is an arbitrary smooth function. It corresponds to the function $g$ introduced in section \ref{sec:algebra}.

One can find the relation between $k$ and $\tilde k$ by considering the  complexified charge $\textbf{Q}_{\tau,\rho} = Q^E_{\tau,\rho} + i Q^M_{\tau,\rho}$.
Demanding that the charges $\textbf{Q}_{\tau,\rho}$ form a closed algebra in the absence of singularities in the parameters
	fixes $\tilde k$ in terms of $k$ to be $\tilde k^2 = -\frac{k^2}{4\lambda} = k^2M^2$.
The algebra of complexified charge with a pole $f=\frac{1}{z-w}$ reads
\begin{align}
	\{
		\textbf{Q}_{\tau,\rho},
		\textbf{Q}_{\tau',\rho'}
	\}
	&=
		\textbf{Q}_{2\tau'',2\rho''}
		+ \left.\frac{1}{2}(D^z D_z^2 f')\right|_{z=w}
	.
\end{align}

\section{Discussions}\label{sec:discussions}
We have explored the consequences of allowing the function parametrizing standard and dual supertranslations on the black hole horizon to have singularities.
The chief outcome is that the algebra of charges develops a central term. What is novel is that we then showed that this anomaly can be cancelled by postulating a holographic Chern-Simons theory on the horizon.
It was noted that both the gravitational case and the electromagnetic one are analogous in this regard.
It is fortunate that the Chern-Simons is a topological theory as it is metric independent.
There are two nice properties that follow. The first is that since the horizon is a null surface,
 the metric is degenerate there. One cannot invert the metric. Had the theory been metric
 dependent, as most are, it would have been impossible to formulate a theory that is restricted to the
 null surface. The second also follows from being metric independent. The energy-momentum tensor
 of a theory is given by varying the action with respect to the metric. Therefore, in the 
 Chern-Simons case, the energy-momentum tensor vanishes and the holographic theory does not
 disturb the black hole geometry.
 
While we have shown that an $\mathrm{SL}(2,\mathbb C)$ Chern-Simons theory cancels the central term, we have not shown that this is the unique theory capable of this cancellation.
There may exist other (arguably more obscure) topological field theories that could accomplish this.
If they exist, then it would be very interesting to see explicit examples of such theories, as they would share properties that are inherent to the structure of the black hole horizon.

It is worth noting that, in the complexified Chern-Simons theory which incorporates both the electric and magnetic actions, the level $\tilde k$ of the magnetic action becomes quantized \cite{Witten:1989ip,Carlip:1994ap}.
For us this is not relevant, since our Chern-Simons theory is just that of the electric action; the magnetic action is only used to compute the dual charge.

 The addition of a holographic Chern-Simons theory on the horizon in conjunction with the soft hair makes the structure at the horizon much more complex. The implications for the nature of black hole microstates will be explored in a future publication.


\begin{acknowledgments}
MJP acknowledges funding from the Science and Technology Facilities Council (STFC) Consolidated Grant ST/T000686/1 ``Amplitudes, Strings and duality''.
MJP would also like to thank the UK STFC for financial support under grant ST/L000415/1.
No new data were generated or analysed during this study.
The work of SC is supported by the European Research Council (ERC) under the European Union’s Horizon 2020 research and innovation programme (grant agreement No 852386).
SC also acknowledges financial support from the Samsung Scholarship.
\end{acknowledgments}

\bibliography{qlett_5-7.bib}

\end{document}